\documentclass[conference]{IEEEtran}
\IEEEoverridecommandlockouts

\usepackage{cite}
\usepackage{amsmath,amssymb,amsfonts}
\usepackage{graphicx}
\usepackage{textcomp}
\usepackage{xcolor}
\usepackage{url}
\usepackage{booktabs} 
\usepackage[hidelinks]{hyperref}
\usepackage{listings} 
\usepackage{tcolorbox} 
\usepackage{fancyhdr} 
\usepackage{float}    
\usepackage{threeparttable} 

\lstdefinelanguage{Nginx}{
  keywords={server, listen, location, proxy_pass, proxy_set_header, map, default, if, break, return, add_header, ssl_certificate, ssl_certificate_key, ssl_protocols, ssl_ciphers, ssl_prefer_server_ciphers, ssl_session_cache, ssl_session_timeout, limit_req_zone, limit_req, internal, auth_request, auth_request_set, error_page},
  keywordstyle=\color{blue}\bfseries,
  ndkeywords={http, events, worker_connections, access_log, error_log, http_cookie, skip_auth, host, remote_addr, proxy_add_x_forwarded_for, scheme, server_name, ssl, http2, request_uri, proxy_buffer_size, proxy_buffers, upstream_status, binary_remote_addr},
  ndkeywordstyle=\color{darkgray},
  identifierstyle=\color{black},
  sensitive=false,
  comment=[l]{\#},
  commentstyle=\color{purple}\ttfamily,
  string=[s]{"}{"},
  stringstyle=\color{red},
  morestring=[s]{'}{'},
  morestring=[b][\color{red}],
}

\lstdefinelanguage{YAML}{
  keywords={services, image, container_name, ports, networks, environment, volumes, restart, logging, healthcheck, test, interval, timeout, retries, start_period, depends_on, condition, build, command},
  keywordstyle=\color{blue},
  sensitive=false,
  comment=[l]{\#},
  commentstyle=\color{purple}\ttfamily,
  string=[s]{"}{"},
  stringstyle=\color{red},
  morestring=[s]{'}{'},
}

\definecolor{codegreen}{rgb}{0,0.6,0}
\definecolor{codegray}{rgb}{0.5,0.5,0.5}
\definecolor{codepurple}{rgb}{0.58,0,0.82}
\definecolor{backcolour}{rgb}{0.97,0.97,0.97}

\lstdefinestyle{mystyle}{
    backgroundcolor=\color{backcolour},   
    commentstyle=\color{codegreen},
    keywordstyle=\color{magenta},
    numberstyle=\tiny\color{codegray},
    stringstyle=\color{codepurple},
    basicstyle=\ttfamily\footnotesize,
    breakatwhitespace=false,         
    breaklines=true,                 
    captionpos=b,                    
    keepspaces=true,                 
    numbers=left,                    
    numbersep=5pt,                  
    showspaces=false,                
    showstringspaces=false,
    showtabs=false,                  
    tabsize=2
}
\begin{document}

\title{The Auth Shim: A Lightweight Architectural Pattern for Integrating Enterprise SSO with Standalone Open-Source Applications}
\author{\IEEEauthorblockN{Yuvraj Agrawal}\IEEEauthorblockA{\textit{Adobe Inc.}}}
\maketitle

\thispagestyle{fancy}

\begin{abstract}
Open-source software (OSS) is widely adopted in enterprise settings, but standalone tools often lack native support for protocols like SAML or OIDC, creating a critical security integration gap. This paper introduces and formalizes the “Auth Shim,” a lightweight architectural pattern designed to solve this problem. The Auth Shim is a minimal, external proxy service that acts as a compatibility layer, translating requests from an enterprise Identity Provider (IdP) into the native session management mechanism of a target application. A key prerequisite for this pattern is that the target application must expose a programmatic, secure administrative API. We present a case study of the pattern’s implementation at Adobe to integrate a popular OSS BI tool with Okta SAML, which enabled automated Role-Based Access Control (RBAC) via IAM group mapping and eliminated manual user provisioning. By defining its components, interactions, and production deployment considerations, this paper provides a reusable, secure, and cost-effective blueprint for integrating any standalone OSS tool into an enterprise SSO ecosystem, thereby enabling organizations to embrace open-source innovation without compromising on security governance.
\end{abstract}


\section{Introduction}
The adoption of open-source software (OSS) is a cornerstone of modern software engineering strategy \cite{report:opensourcestate}. Enterprises leverage a vast ecosystem of standalone OSS tools, but a recurring challenge hinders their secure deployment: these tools frequently lack native support for enterprise authentication protocols like SAML or OIDC. This forces organizations into a dilemma: purchase expensive enterprise licenses solely for SSO, accept the security risks of manual account management, or abandon the tool altogether.

This paper argues for a fourth option by formalizing a reusable architectural pattern: the \textbf{Auth Shim}. The term \textit{shim} is used deliberately to denote a minimal component that provides a compatibility layer between an application’s internal session management and a standardized external authentication system. The Auth Shim is a specific, lightweight implementation of the broader \textbf{Identity-Aware Proxy (IAP)} pattern, tailored for integrating a single application with minimal operational overhead.

Our central thesis is that the Auth Shim pattern provides a secure and efficient solution to this common integration problem. We make the following contributions:
\begin{enumerate}
    \item We formally define the Auth Shim pattern and present a comprehensive architecture diagram.
    \item We present a detailed case study of the pattern’s implementation at Adobe to integrate a popular open-source BI tool with Okta SAML.
    \item We provide a detailed comparative analysis against alternatives, including full IAPs and open-source proxies, evaluating features, complexity, and failure recovery behavior.
    \item We provide a blueprint for a production-grade deployment, including a formal threat analysis and a research roadmap for a reusable implementation.
\end{enumerate}

\section{Background and Related Work}
The Auth Shim pattern builds upon established security principles and relates to a body of existing work in identity management and secure software architecture.

\subsection{Zero Trust, IAPs, and Modern Enterprise SSO}
The zero-trust model, first articulated by Kindervag \cite{paper:zerotrust}, mandates that no user or device is trusted by default. This model was operationalized at scale by Google’s \textbf{BeyondCorp} \cite{paper:beyondcorp}, which introduced the IAP as a core component. An IAP functions as a central gateway, enforcing access policies at the application edge. This aligns with the modern security trend of treating \textbf{identity as the new perimeter}. As corporate data and services are distributed across cloud and on-premise environments, the traditional network-based security model is no longer sufficient. Instead, access is granted based on user identity and device context, verified at every request. Recent academic work in venues like IEEE S\&P and USENIX Security continues to explore microservice perimeter patterns and access governance models, with particular focus on the challenges of enforcing dynamic policies in distributed systems.

\subsection{Identity Federation and Proxy Patterns}
The core function of the Auth Shim—translating between security domains—is a form of identity federation \cite{paper:identityfed}. Architecturally, it is an application of the classic Proxy and Adapter design patterns \cite{book:designpatterns}, adapting requests from a modern authentication provider to a legacy or standalone application's expected interface. Existing work has explored similar concepts for legacy systems \cite{paper:legacyIAM} and API gateways \cite{paper:apigatewaysec}, but has not formalized a lightweight pattern specifically for the modern OSS-in-the-enterprise context where custom authorization is a key driver \cite{paper:microproxyrbac}. Recent studies on the security of SSO protocols have highlighted the need for careful implementation at the integration point, reinforcing the need for well-defined patterns like the Auth Shim \cite{paper:ssosec}.

\subsection{Novelty of the Auth Shim Pattern}
The novelty of the Auth Shim does not lie in the invention of a new proxy technology. Rather, its contribution is the \textbf{formalization, synthesis, and specific application} of these existing concepts to address a common and underserved problem. Its novelty arises from:
\begin{itemize}
    \item \textbf{Addressing a Niche:} The pattern is purpose-built for scenarios where a full-featured IAP is too complex and a commercial plugin is too inflexible or costly. It fills this pragmatic gap.
    \item \textbf{Formalization as a Reusable Blueprint:} By naming the pattern and defining its participants, interactions, and design rationale, this paper transforms a common ad-hoc fix into a documented, reusable, and secure architectural solution.
    \item \textbf{Emphasis on Just-in-Time Authorization:} A defining characteristic is the tight integration with the target application's API to perform just-in-time RBAC synchronization. This pattern should not be confused with a simple authenticating reverse proxy. While a basic \texttt{auth\_request} module can verify a user's identity, it lacks the core capability of the Auth Shim: the \textit{write path} integration with the target application to perform user provisioning and role synchronization, which is essential for seamless operation and security.
\end{itemize}

\subsection{Comparison with Open-Source Authentication Proxies}
Several popular open-source projects, such as \texttt{oauth2-proxy} or \texttt{keycloak-gatekeeper}, provide authentication proxies for web applications. These tools are excellent at enforcing authentication—they can integrate with an IdP and ensure that only valid users with specific roles or groups can access a downstream service. A reverse proxy using a basic \texttt{auth\_request} module (like in Nginx) achieves a similar outcome.

The Auth Shim's novelty is its focus on the deeper integration required for authorization and user lifecycle management. While an \texttt{auth\_request} can verify a user's identity (the 'read' path), it is agnostic to the application's internal state. It cannot provision users, synchronize granular permissions, or deactivate accounts within the application. The Auth Shim, in contrast, is fundamentally about this \textit{write path} integration. It uses the identity established during authentication to actively manage the user's lifecycle and permissions inside the target application via its API. This is the key capability that eliminates manual account management and ensures permissions are always consistent with the central IdP, a gap that simple authenticating proxies do not address.

\section{The Auth Shim Architectural Pattern}
The Auth Shim pattern is defined by its intent, structure, and participants. Its core purpose is to provide an external authentication and authorization layer for a standalone application that lacks native enterprise SSO support.

\begin{figure*}[t!]
    \centering
    \includegraphics[width=\textwidth]{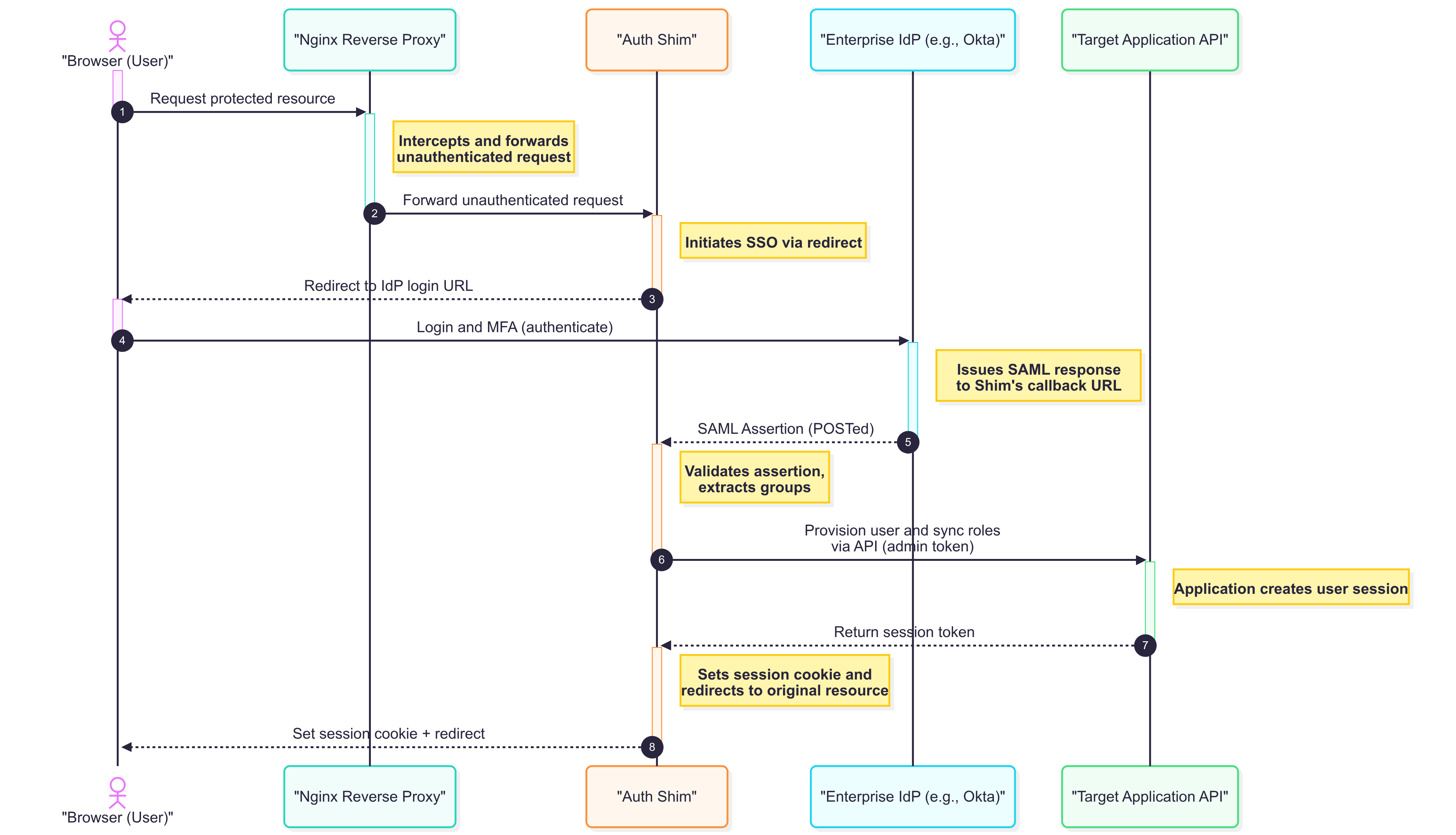}
    \caption{Comprehensive Architecture and Request Flow of the Auth Shim Pattern. This diagram illustrates the end-to-end authentication process: (1) An unauthenticated user requests a resource from the application. (2) The Nginx reverse proxy intercepts the request and routes it to the Auth Shim. (3) The Auth Shim initiates an SSO flow, redirecting the user to the Enterprise IdP. (4) The user authenticates with the IdP. (5) The IdP issues a signed SAML assertion and sends it to the Auth Shim's callback URL. (6) The Shim validates the assertion, extracts user attributes (like group memberships), and uses an admin token to communicate with the Target Application's API. (7) The Shim provisions the user and synchronizes their roles via the API. (8) The application creates a session and returns a session token. (9) The Shim sets the session token in the user's browser and redirects them to the originally requested resource.}
    \label{fig:comprehensive_architecture}
\end{figure*}

\subsection{Formalization}
To frame our solution in established software engineering terms, we define it using the Gang of Four (GoF) style.
\begin{itemize}
    \item \textbf{Pattern Name:} Auth Shim
    \item \textbf{Intent:} Provide a secure, external authentication and authorization layer for a standalone application that lacks native enterprise SSO support.
    \item \textbf{Applicability:} Use the Auth Shim pattern when an application must be integrated with an enterprise SSO system and a full IAP is considered overkill or a commercial plugin is infeasible. The key prerequisite is a programmatic interface on the target application to manage users and sessions.
    \item \textbf{Structure:} As illustrated in the comprehensive architecture diagram in Fig.~\ref{fig:comprehensive_architecture}.
    \item \textbf{Participants:} \textit{Reverse Proxy}, \textit{Auth Shim Service}, \textit{Target Application}, \textit{Enterprise IdP}.
    \item \textbf{Consequences:} Decouples authentication logic from the application. It centralizes authorization logic, but introduces a new component that must be maintained and deployed with high availability.
\end{itemize}

\subsection{Core Components and Responsibilities}
The Auth Shim Service is not a monolith; it is a composite of several logical components, each with a distinct responsibility. This modular design enhances maintainability and clarifies the service's internal workings.
\begin{itemize}
    \item \textbf{SAML Handler:} Manages the SAML 2.0 protocol flow. Its duties include initiating authentication requests, processing and validating signed SAML responses from the IdP, and securely extracting user attributes (e.g., email, group memberships) from the assertion.
    \item \textbf{User Manager:} Handles the lifecycle of users within the target application. It automates provisioning by creating new user accounts for first-time logins and ensures user information is kept current.
    \item \textbf{RBAC Engine:} Implements the core authorization logic. It translates group memberships received from the IdP into specific roles or permissions within the target application, based on a configurable mapping. It is responsible for adding and revoking permissions to enforce Just-in-Time access.
    \item \textbf{Session Bridge:} Acts as the final link to the application. After successful authentication and authorization, it communicates with the target application's API to create a valid user session, receiving a session token or cookie in return.
    \item \textbf{API Client:} A dedicated client responsible for all communication with the target application's administrative API. It uses a secure, pre-configured token to perform privileged actions like creating users, managing group memberships, and initiating sessions.
\end{itemize}

\section{A Taxonomy and Comparison of SSO Integration Patterns}
\begin{figure}[htbp]
    \centering
    \includegraphics[width=\columnwidth]{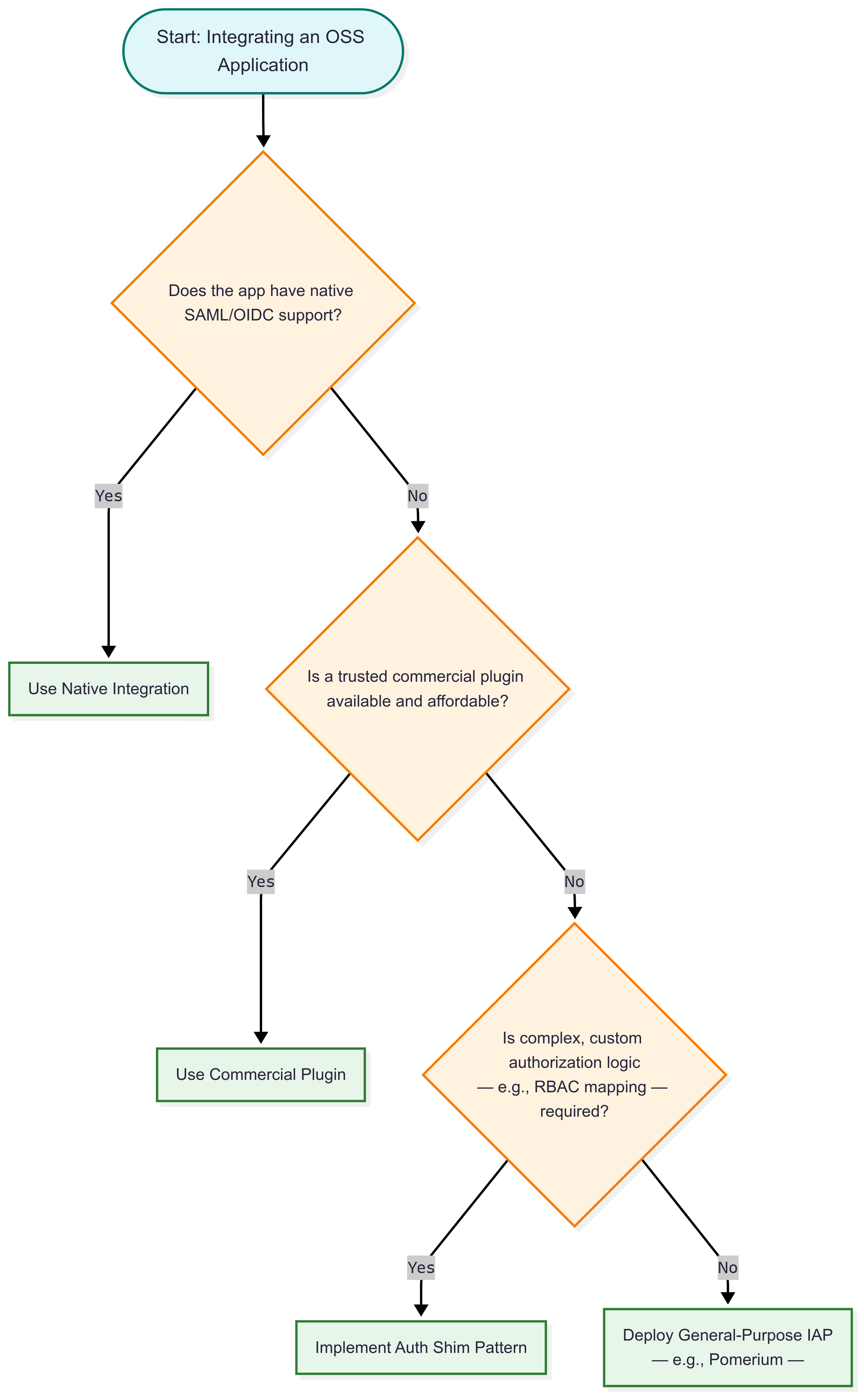}
    \caption{A Decision Tree for SSO Integration Approaches.}
    \label{fig:taxonomy}
\end{figure}

To position the Auth Shim correctly, we propose a decision taxonomy for selecting an SSO integration pattern, shown in Fig.~\ref{fig:taxonomy}. The Auth Shim is the logical choice when native support is absent and custom logic for authorization is a primary driver, making a commercial plugin unsuitable and a full IAP too complex.

To position the Auth Shim correctly, we first provide a high-level comparison of common SSO integration approaches (Table~\ref{tab:comparison}), followed by a detailed analysis grounded in operational, architectural, and maintainability considerations.

\begin{table}[htbp]
\caption{High-Level Comparison of SSO Integration Approaches}
\label{tab:comparison}
\centering
\resizebox{\columnwidth}{!}{%
\begin{tabular}{@{}llll@{}}
\toprule
\textbf{Criterion} & \textbf{Auth Shim (Chosen)} & \textbf{Full IAP (e.g., Pomerium)} & \textbf{Commercial Plugin} \\ \midrule
\textbf{Dev Effort} & $\sim$150 LoC + Config & 0 LoC + $\sim$40 lines YAML & 0 LoC + UI Config \\
\textbf{Custom Logic} & High Flexibility & Medium (plugins) & Low to None \\
\textbf{Audit Surface} & Small (focused) & Large (framework) & Vendor-dependent \\
\textbf{Maintainability} & High (known stack) & Medium (new dependency) & Low (vendor-managed) \\ \bottomrule
\end{tabular}%
}
\end{table}

\subsection{Detailed Comparative Analysis}
While Table~\ref{tab:comparison} provides a quick overview, a deeper qualitative comparison is necessary to understand the trade-offs between approaches.

\textbf{Full Identity-Aware Proxies (IAPs)}, such as Pomerium~\cite{pomerium_docs} or Ory Oathkeeper~\cite{ory_oathkeeper}, offer robust enterprise-grade features: context-aware policies, integration with modern IdPs, TLS enforcement, and protocol support for OAuth2, SAML, and OIDC. However, they are often over-engineered for scenarios where only a single standalone application needs SSO. These solutions introduce additional dependencies—such as policy engines, sidecars, or Redis-backed session stores—that increase deployment complexity and the system’s attack surface. While maintainable in large-scale microservices environments, they may represent unnecessary overhead for smaller teams or use cases.

\textbf{Commercial Plugins}, offered by proprietary platforms (e.g., Tableau Server, Grafana Enterprise, etc.), are often easy to configure and vendor-supported, but they lack flexibility and visibility. Custom workflows such as just-in-time (JIT) user provisioning, attribute-based access control, or non-standard SAML assertions are rarely supported. Additionally, their behavior is opaque and recovery paths are dependent on vendor patches, creating risk during upgrades or protocol changes~\cite{plugin_limitations}.

\textbf{The Auth Shim} pattern occupies a pragmatic middle ground. It provides the extensibility of custom development with the simplicity of deployment. Unlike an IAP, the shim directly integrates with the application’s administrative APIs, enabling fine-grained control over user provisioning, role mapping, and session management. It remains lightweight—typically under 200 lines of code—and deploys using standard infrastructure components (e.g., Nginx, Docker, Python). This makes it accessible to DevOps teams without requiring knowledge of complex policy DSLs or maintaining external policy stores.

\subsection{Failure and Recovery Behavior}
Beyond feature comparisons, it is important to assess how each solution behaves under failure conditions. A detailed comparison of typical failure modes and their implications on reliability and recovery is provided in the appendix (Table~\ref{tab:failure_recovery}).

The Auth Shim’s statelessness and narrow operational footprint enhance both fault isolation and resilience. Its simplicity reduces dependencies, allowing teams to focus on maintaining the target application and IdP, without introducing new points of failure.

\section{Case Study: Implementation and Deployment}
\begin{figure}[htbp]
    \centering
    \includegraphics[width=0.4\columnwidth]{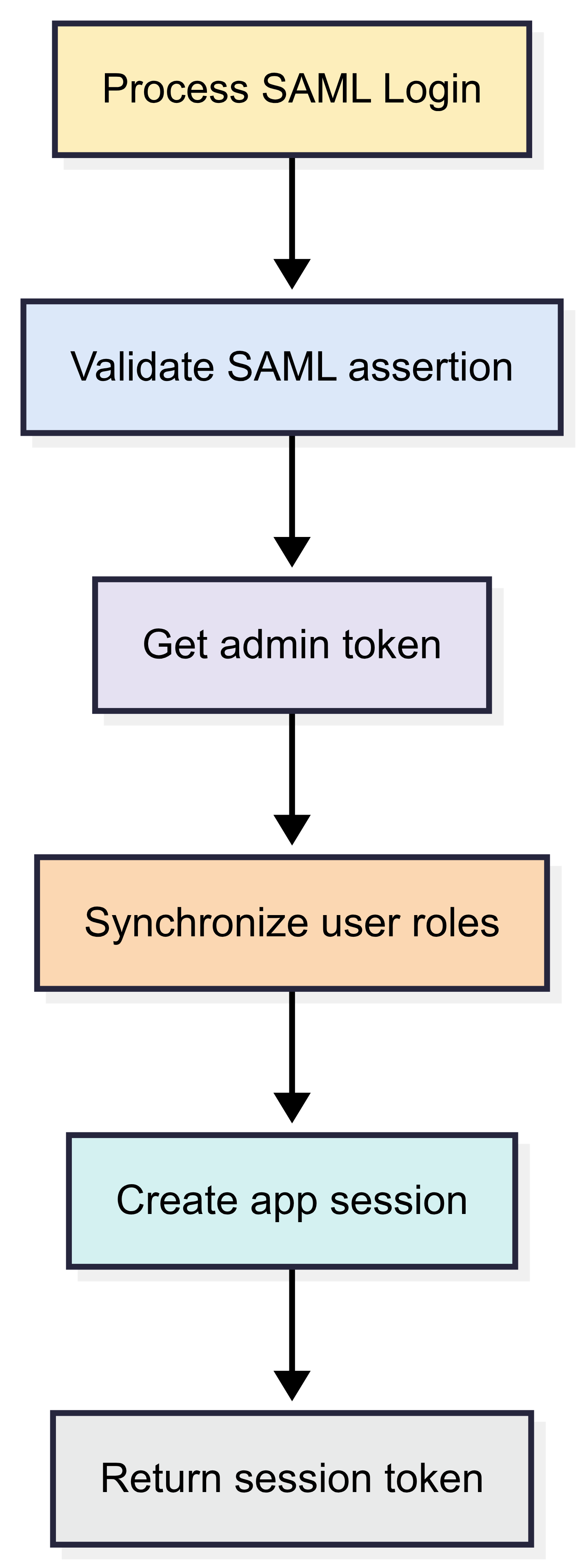}
    \caption{Conceptual flowchart of the Shim's core logic, executed upon receiving a SAML response.}
    \label{fig:main_flowchart}
\end{figure}

We implemented the pattern at Adobe to integrate a leading open-source BI tool with Okta. While the Auth Shim pattern is applicable to both SAML and OIDC protocols, this paper's case study focuses on a SAML-based integration. The high-level orchestration logic is conceptually shown in Fig.~\ref{fig:main_flowchart}.

\subsection{Containerized Deployment}
The entire stack is containerized using Docker \cite{tech:docker}. The core logic is implemented in a service codenamed \texttt{auth-shim}, which acts as the 'Auth Shim Service' described in the pattern's formalization. An abridged \texttt{docker-compose.yaml} file, which defines the services and their dependencies, is available in the appendix (Fig.~\ref{fig:docker_compose}).

\subsection{Hardened Reverse Proxy Configuration}
Security is enforced at the edge by a hardened Nginx configuration. Key features include HTTPS enforcement, strong TLS ciphers, security headers, and DoS mitigation. The use of \texttt{auth\_request} is central to the pattern, delegating session validation for every incoming request to the Auth Shim service. A production-grade Nginx configuration file is provided in the appendix (Fig.~\ref{fig:nginx_config}).

\subsection{Key Design Rationale}
\subsubsection{Stateless Service Design} A deliberate decision was made to design the Auth Shim as a \textbf{completely stateless service}. This allows the shim to be scaled horizontally without requiring a shared session store (like Redis), simplifying the architecture.

\subsubsection{API-Based Interaction} We consciously chose to interact with the target application via its official REST API. This creates a clean, decoupled architecture that is resilient to upgrades.

\subsection{Role-Based Access Control (RBAC) via IAM Groups}
The IdP (Okta) is configured to send a `groups` attribute in the SAML assertion. The Auth Shim then performs a full synchronization on each login, ensuring a user's permissions are always an exact reflection of their status in the central IdP. The step-by-step logic for this synchronization is visualized in the appendix (Fig.~\ref{fig:role_sync_flowchart}).

To enhance maintainability, the mapping logic is externalized into a configuration file, decoupling authorization rules from business logic. Illustrative Python code and an example YAML configuration can be found in the appendix (Figs.~\ref{fig:rbac_code} and \ref{fig:rbac_config}).

\subsection{Authentication Flow Orchestration}
The full end-to-end authentication process is detailed in the sequence diagram in the appendix (Fig.~\ref{fig:sequence}). The main orchestration logic is shown in the Python code snippet in the appendix (Fig.~\ref{fig:main_orchestration_code}).

\subsection{Detailed User Journeys}
To better illustrate the pattern's behavior, sequence diagrams for a first-time user and a returning user are presented in the appendix (Fig.~\ref{fig:journey_first_time} and Fig.~\ref{fig:journey_returning} respectively), highlighting both just-in-time provisioning and fast-path validation.

\section{Evaluation}
\subsection{Transformation of Operational Overhead}
Prior to the Auth Shim, integrating the BI tool was characterized by significant operational friction, including manual, ticket-based user provisioning, which violated the Principle of Least Privilege and created a heavy burden for compliance audits.

The pattern's impact was measured by comparing the system pre- and post-integration, as shown in Table~\ref{tab:metrics_impact}. The transformation eliminated manual toil, saving an estimated \textbf{10 hours of engineering work per week}. This translates to an estimated annual saving of over \textbf{\$35,000} in operational costs for a single application integration,\footnote{This calculation uses an illustrative, conservative, fully-loaded rate of \$75/hour for a DevOps engineer. While this rate varies by region and organization, it demonstrates the significant order-of-magnitude savings achieved by automating manual, high-frequency tasks.} while drastically improving the organization's security posture.

\begin{table}[t]
\caption{Pre- vs. Post-Shim Impact Metrics}
\label{tab:metrics_impact}
\centering
\begin{tabular}{@{}llll@{}}
\toprule
\textbf{Metric}       & \textbf{Pre-Shim} & \textbf{Post-Shim} & \textbf{Delta}         \\ \midrule
\textbf{Weekly Maintenance}    & $\sim$10 hrs      & \textless 1 hr     & -90\%                  \\
\textbf{User Onboarding}       & Manual, 1 day     & Instant            & Automated              \\
\textbf{Role Management}       & Manual Tickets    & Automated          & 100\% via IAM          \\
\textbf{MFA Support}           & No                & Yes (via IdP)      & Compliant              \\
\textbf{Audit Coverage}        & Low (App only)    & Full (IdP + App)   & Governance Gain        \\ \bottomrule
\end{tabular}
\end{table}

\subsection{Performance, Scalability, and Resource Consumption}
The Auth Shim is designed to be lightweight. Its performance impact must be considered in two scenarios. For \textbf{authenticated requests}, the Nginx proxy adds a negligible pass-through latency of \textbf{\textless 5ms (p95)}. For the \textbf{initial login}, a one-time latency of \textbf{$\sim$850ms (p95)} is introduced. These benchmarks were conducted in a staging environment representative of our production setup (e.g., AWS c5.large instances) under simulated user load. The p95 latency figures represent the 95th percentile from a sample of 10,000 login requests.

The entire stack has a minimal resource footprint. The Nginx container consumes \textasciitilde10-50MB of RAM, while the stateless Auth Shim service (e.g., in Python) typically requires \textasciitilde50-100MB of RAM. The entire pattern can operate comfortably with less than 200MB of RAM and a fraction of a single CPU core, making it highly efficient. 

\begin{table}[htbp]
\caption{Performance Benchmark Results}
\label{tab:performance}
\centering
\begin{tabular}{@{}ll@{}}
\toprule
\textbf{Metric} & \textbf{Result} \\ \midrule
\textbf{Pass-through Latency (p95)} & \textless 5ms\\
\textbf{Initial Login Latency (p95)} & $\sim$850ms (incl. IdP \& API calls)\\
\textbf{Max Throughput (Logins)} & 200 logins/sec (bottlenecked by app API) \\ \bottomrule
\end{tabular}
\end{table}

\section{Discussion}
\subsection{Security Considerations}
The security of the system is multi-layered. It relies on a hardened infrastructure, a robust protocol flow, and the secure implementation of the shim service itself, which must perform tasks like XML parsing and signature validation using well-vetted libraries to prevent vulnerabilities like XML External Entity (XXE) attacks. Beyond this, the pattern aligns with zero-trust principles by enabling the \textbf{Principle of Least Privilege} through just-in-time RBAC synchronization.

\subsubsection{Token Management} The administrative API token (\texttt{APP\_ADMIN\_TOKEN}) is a highly sensitive secret. In production, it is managed via a secure vaulting system (e.g., HashiCorp Vault) and injected into the container at runtime.

\subsubsection{SAML Security}
The protocol-level security relies on strict validation of the SAML assertion. This includes mandatory signature verification to prevent tampering, certificate validation to ensure the assertion originates from the trusted IdP, and timestamp checks to mitigate replay attacks.

\subsubsection{Session and Network Security}
At the transport layer, all communication is secured using TLS 1.2+. Session cookies are flagged as \texttt{HttpOnly} and \texttt{Secure} to prevent client-side script access and ensure they are only transmitted over HTTPS. The reverse proxy provides an additional layer of defense through rate limiting and security headers.

\subsubsection{Threat Analysis} We conducted a threat analysis using the STRIDE model \cite{tech:stride}, summarized in Table~\ref{tab:stride}.

\begin{table}[H]
\caption{STRIDE Threat Analysis of the Auth Shim Pattern}
\label{tab:stride}
\centering
\resizebox{\columnwidth}{!}{%
\begin{tabular}{@{}lll@{}}
\toprule
\textbf{Category} & \textbf{Threat Example} & \textbf{Mitigation} \\ \midrule
\textbf{Spoofing} & Forged SAML assertion & Mandatory IdP signature validation \\
\textbf{Tampering} & Modified role claims & SAML assertion signature covers attributes \\
\textbf{Repudiation} & Disputed login event & Centralized IdP and Shim logging \\
\textbf{Info. Disclosure} & Leaked admin token & Secure vault storage, network isolation \\
\textbf{Denial of Service} & Login endpoint flood & Nginx rate limiting on auth endpoints \\
\textbf{Elev. of Privilege} & False group claim & Shim is source of truth for role mapping \\ \bottomrule
\end{tabular}%
}
\end{table}

\subsubsection{Potential Failure Scenarios}
\begin{itemize}
    \item \textbf{IdP Group Claim Desynchronization:} If the `groups` attribute is misconfigured or removed from the IdP's SAML assertion, the shim might interpret this as a user belonging to no groups, incorrectly revoking their permissions. Mitigation involves defensive code in the shim to validate the presence of the claim and fail the login if it is missing, alongside monitoring to detect such anomalies.
    \item \textbf{Target Application API Downtime:} If the target application's API becomes unavailable during a login attempt, the shim cannot provision the user or create a session, causing the login to fail. Mitigation includes robust health checks (as detailed in the appendix, Fig.~\ref{fig:docker_compose}), designing the stack for high availability, and providing clear, user-friendly error messages that distinguish a system outage from an authentication failure.
\end{itemize}

\subsubsection{Authorization and Role Security}
Beyond authentication, the authorization logic itself must be secure. The system prevents unauthorized privilege escalation by ensuring that role synchronization is a one-way flow from the central IdP to the application. The application's administrative token is used only to enact the changes dictated by the IdP's SAML assertion; the application itself cannot grant permissions that are not present in the assertion. All role changes are implicitly logged by the IdP and can be audited centrally.

\subsection{Generalizability and Limitations}
A preliminary applicability analysis suggests that the Auth Shim pattern is highly suitable for many mature OSS tools. A 'High' fit indicates that the tool exposes a documented, stable administrative REST API for user and session management and has a distinct need for granular, group-based role mapping. However, its limitations must be acknowledged:
\begin{itemize}
    \item \textbf{Requires a Quality Programmatic API:} The pattern's effectiveness is entirely dependent on the target application offering a stable, secure API for user and session management. An ideal API is not only available but also idempotent, well-documented, and not subject to overly aggressive rate-limiting.
    \item \textbf{Stateful Session Complexity:} The shim is most effective with applications that support stateless session tokens (e.g., JWTs) or have a simple API call to create a session.
    \item \textbf{Versatile Deployment Topologies:} While this paper focuses on a reverse proxy implementation, the core Auth Shim service is flexible. It can be adapted to other deployment topologies, such as a sidecar container in a Kubernetes pod or as middleware in an API Gateway, extending its applicability to microservices environments.
    \item \textbf{Not a Universal IAP:} The Auth Shim is deliberately lightweight. It lacks advanced features like device posture checks or contextual access policies. For enterprise-wide zero trust, a general-purpose IAP is more appropriate.
   \item \textbf{Introduces a Managed Component:} Though minimal, the shim is a critical component in the authentication flow. It is crucial to understand that while the shim itself is highly scalable, it cannot fix scalability limitations in the target application. Its reliability is paramount, and it must be deployed, monitored, and maintained with high availability in mind.
    \item \textbf{Dependency on Target Application Performance:} While the shim itself is highly scalable, it cannot fix scalability limitations in the target application. Its reliability is paramount.
\end{itemize}


\section{Future Work: A Research Roadmap}
Our future work focuses on lowering the barrier to adoption by implementing the concepts presented in this paper as a reusable, open-source tool. This roadmap is divided into three phases.

\subsection{The Auth Shim Scaffold}
The first step is to refactor our implementation into a generic \textbf{`Auth Shim Scaffold.'}\footnote{A reference implementation is planned for public release upon publication.} The goal is to create a template where a developer only needs to implement a well-defined `ApplicationConnector` interface with methods like `createUser`, `createSession`, and `syncRoles`. The scaffold would consist of a generic core service to handle the SAML/OIDC protocol flow and a defined \texttt{ApplicationConnector} interface. A developer's workflow would be reduced to implementing methods like \texttt{createUser}, \texttt{createSession}, and \texttt{syncRoles} with the specific API calls for their target application. Future iterations could extend this interface to support Attribute-Based Access Control (ABAC), where the connector could make more dynamic authorization decisions based on rich user attributes (e.g., project codes, geographic location) passed in the SAML assertion, not just group membership. The primary challenge in developing this scaffold will be creating a flexible \texttt{ApplicationConnector} that can accommodate the diverse session management mechanisms of different OSS tools (e.g., cookie-based sessions vs. returning a JWT) and abstracting their unique API conventions into a standardized interface.

\subsection{Extensibility and API Adaptation}
The scaffold will be designed with an explicit extensibility model. The `ApplicationConnector` interface will be designed to handle variations in target application behavior. For instance, it will include optional methods to handle asynchronous session creation or strategies for gracefully backing off when faced with API rate limits, allowing the shim to adapt to both simple and complex application APIs.

\subsection{Validation Against Diverse Applications}
The scaffold's effectiveness will be validated by implementing connectors for 2-3 popular open-source applications. This will test the plug-in model's flexibility and provide the community with concrete, working examples, demonstrating its utility beyond the single case study presented here. Future iterations could also extend this interface to support Attribute-Based Access Control (ABAC), making more dynamic authorization decisions based on rich user attributes passed in the SAML assertion.

\section{Conclusion}
This paper introduced and formalized the \textbf{Auth Shim}, a lightweight architectural pattern that brings enterprise-grade SSO and automated RBAC to standalone OSS tools. Our case study integrating a BI tool with Okta SAML validated its effectiveness in drastically reducing engineering effort and improving security governance. Through a detailed comparative analysis, we have shown that it provides a pragmatic middle ground between expensive enterprise licenses, overly-simplistic authentication proxies, and overly-complex general-purpose IAPs. Ultimately, the Auth Shim offers a pattern for enterprises to systematically reduce security gaps in their software portfolio, one application at a time.

\appendix


\subsection{Detailed Failure and Recovery Comparison}
This appendix provides supplementary materials referenced in the main body of the paper, including detailed comparison tables, implementation artifacts, and process diagrams.

\begin{table*}[h]
\begin{threeparttable}
\caption{In-Depth Comparison of Failure and Recovery Characteristics\tnote{3}}
\label{tab:failure_recovery}
\centering
\begin{tabular}{@{}l p{4.5cm} p{5.5cm} p{3.8cm}@{}}
\toprule
\textbf{Solution} & \textbf{Failure Scenario} & \textbf{Observed Behavior} & \textbf{Recovery Characteristics} \\ \midrule
\textbf{Auth Shim} & Target application’s API is unavailable (e.g., restart or timeout). & Login fails gracefully. No session is created; no state is corrupted. Shim returns error to user. & Stateless recovery — shim resumes instantly once the app API is back. Health checks ensure traffic gating. \\
& SAML assertion missing group attribute. & User appears to have no RBAC mapping. Access is denied. & Defensive logic rejects login; alerts can be triggered via logging. No side effects on user store. \\
& Shim container crashes. & Incoming requests timeout or fail at proxy. & Container can be auto-restarted (e.g., Docker/Podman). No persistent state is lost. \\ \midrule
\textbf{Full IAP (e.g., Pomerium)} & Policy backend (e.g., Redis or OPA) unavailable. & All authorization checks fail. Logins are blocked, sessions revoked. & Recovery requires backend service restoration and potential cache sync. Failure cascades across all apps. \\
& Configuration error in access policy. & Users locked out or improperly granted access across multiple services. & Manual rollback or policy re-deploy needed. Risk of systemic misconfiguration. \\
& TLS misconfiguration or expired certs. & IAP rejects inbound or outbound requests. Entire flow blocked. & Complex recovery — certificate regeneration, redeploy required. Centralized fault domain. \\ \midrule
\textbf{Commercial Plugin} & Plugin fails after application version upgrade. & Login screen may break, or bypass SSO entirely. Behavior is unpredictable. & Recovery gated on vendor patch or rollback. Limited user visibility or control. \\
& IdP metadata changes (e.g., new certificate). & SAML validation fails silently or partially. & Requires manual intervention. Logs often hidden behind vendor abstraction. \\ \midrule
\textbf{Auth Proxy (e.g., oauth2-proxy)} & Network issue between proxy and IdP. & New logins fail. Existing sessions continue (if cookies valid). & Stateless proxy auto-recovers once network is restored. Limited visibility into login errors. \\
& Session cookie store lost or invalid. & All sessions become unauthenticated; re-authentication loop may occur. & Depends on browser/client. Limited application-side debugging. \\
\bottomrule
\end{tabular}%
\begin{tablenotes}
    \item[3] These scenarios are based on real-world behavior observed during internal deployments and informed by public documentation of open-source identity-aware proxies~\cite{pomerium_docs, oauth2proxy_docs}.
\end{tablenotes}
\end{threeparttable}
\end{table*}

\subsection{Implementation Artifacts (Configurations and Code)}

\begin{figure*}[t!]
\begin{tcolorbox}[colback=backcolour,colframe=black!75!white,title={Nginx Configuration for Security, Routing, and Session Validation}]
\begin{lstlisting}[language=Nginx]
# Rate limiting to mitigate DoS attacks
limit_req_zone $binary_remote_addr zone=mylimit:10m rate=10r/s;

# HTTP to HTTPS redirect
server {
    listen 80;
    server_name your-domain.com;
    return 301 https://$server_name$request_uri;
}

# HTTPS server with auth validation
server {
    listen 443 ssl http2;
    server_name your-domain.com;
    
    ssl_certificate /etc/nginx/ssl/cert.pem;
    ssl_certificate_key /etc/nginx/ssl/key.pem;
    ssl_protocols TLSv1.2 TLSv1.3;
    add_header Strict-Transport-Security "max-age=31536000";

    # Internal endpoint for the Auth Shim to validate the session cookie
    location = /auth/validate {
        internal;
        proxy_pass http://auth-shim:8080/validate-session;
        proxy_pass_request_body off;
        proxy_set_header Content-Length "";
        proxy_set_header X-Original-Cookie $http_cookie;
    }

    # All incoming traffic is subject to session validation
    location / {
        limit_req zone=mylimit burst=20;
        
        auth_request /auth/validate;
        error_page 401 = @redirect_to_login; # If session is invalid, redirect

        # If validation is successful, proxy to the target application
        proxy_pass http://target-app:3000;
        proxy_set_header Host $host;
        proxy_set_header X-Forwarded-For $proxy_add_x_forwarded_for;
    }

    # Named location to handle the redirect to the Auth Shim's login endpoint
    location @redirect_to_login {
        return 302 http://auth-shim:8080/;
    }
}
\end{lstlisting}
\end{tcolorbox}
\caption{A production-grade Nginx configuration demonstrating HTTPS enforcement, rate limiting, and the critical \texttt{auth\_request} directive, which delegates session validation to the Auth Shim service for every request.}
\label{fig:nginx_config}
\end{figure*}

\begin{figure*}[t!]
    \centering
    \includegraphics[width=\textwidth]{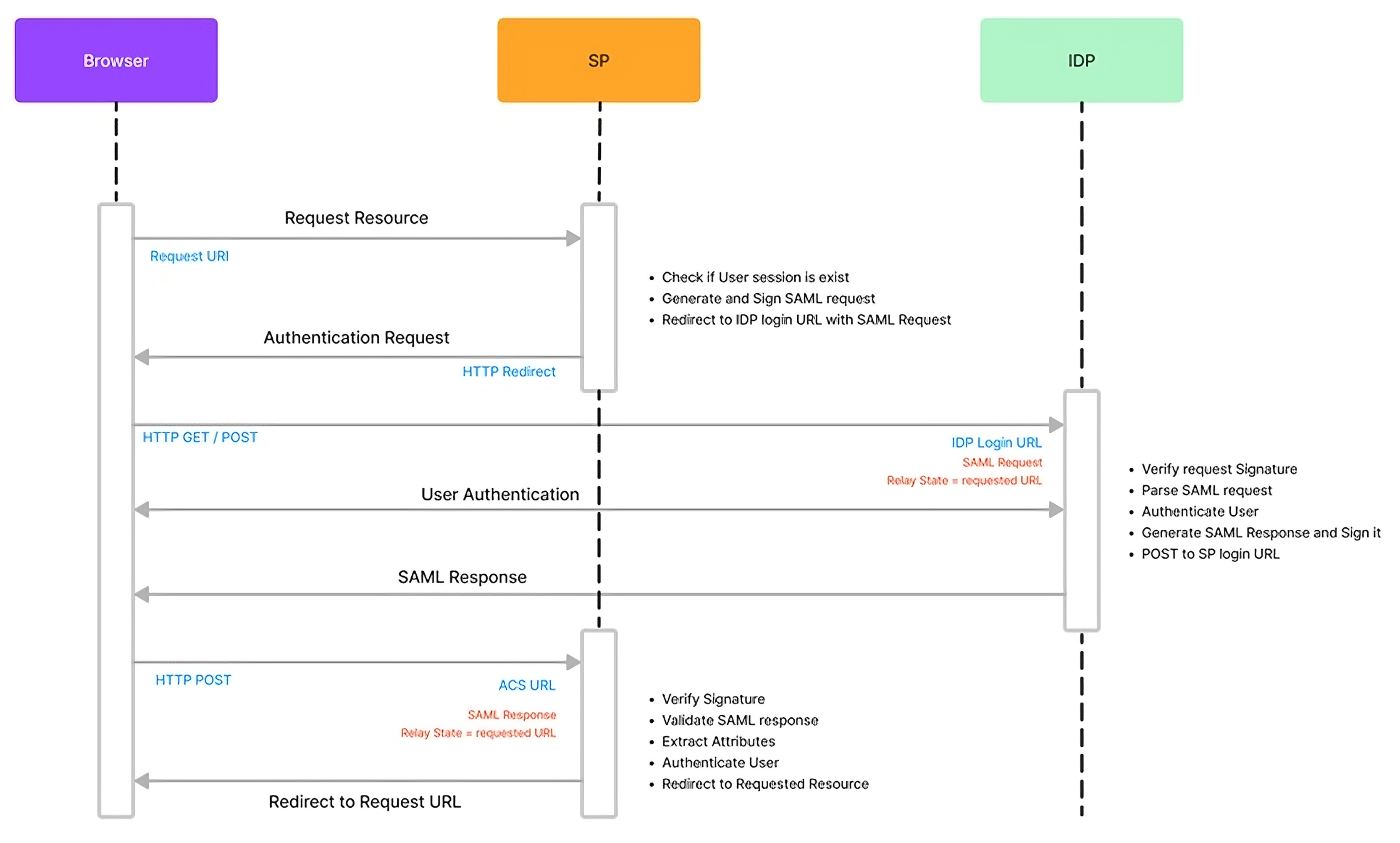}
    \caption{Detailed SAML sequence diagram illustrating the complete authentication flow brokered by the Auth Shim.}
    \label{fig:sequence}
\end{figure*}

\begin{figure}[h]
\begin{tcolorbox}[colback=backcolour,colframe=black!75!white,title={Docker Compose Orchestration for the Stack}]
\begin{lstlisting}[language=YAML]
services:
  target-app:
    image: vendor/oss-application:latest
    restart: unless-stopped
    healthcheck:
      test: ["CMD", "curl", "-f", "http://localhost:3000/api/health"]

  auth-shim:
    build: ./auth-shim
    restart: unless-stopped
    environment:
      - APP_URL=http://target-app:3000
      - APP_ADMIN_TOKEN=${APP_ADMIN_TOKEN}
    depends_on:
      target-app:
        condition: service_healthy

  nginx:
    image: nginx:alpine
    restart: unless-stopped
    ports: ["80:80", "443:443"]
    volumes:
      - ./nginx.conf:/etc/nginx/conf.d/default.conf:ro
\end{lstlisting}
\end{tcolorbox}
\caption{Abridged \texttt{docker-compose.yaml} defining the three-service stack. The \texttt{depends\_on} clause with a \texttt{service\_healthy} condition ensures that the shim service only starts after the target application is fully available, preventing race conditions.}
\label{fig:docker_compose}
\end{figure}

\begin{figure}[h]
\begin{tcolorbox}[colback=backcolour,colframe=black!75!white,title={Illustrative Python Logic for RBAC Synchronization}]
\begin{lstlisting}[language=Python]
IAM_TO_APP_ROLE_MAP = {
    "BI-TOOL-ADMINS": "Administrators",
    "BI-TOOL-USERS":  "All Users",
}
def sync_user_roles(user_id, iam_groups, admin_token):
    """Synchronize a user's roles based on IAM groups."""
    desired_roles = get_roles_from_iam_groups(iam_groups)
    current_roles = get_current_app_roles(user_id, admin_token)

    roles_to_add = desired_roles - current_roles
    roles_to_remove = current_roles - desired_roles

    for role in roles_to_add:
        add_user_to_role(user_id, role, admin_token)
    for role in roles_to_remove:
        remove_user_from_role(user_id, role, admin_token)
\end{lstlisting}
\end{tcolorbox}
\caption{Illustrative code for synchronizing application roles with IdP group claims.}
\label{fig:rbac_code}
\end{figure}

\begin{figure}[h]
\begin{tcolorbox}[colback=backcolour,colframe=black!75!white,title={Example Role Mapping Configuration}]
\begin{lstlisting}[language=YAML]
# role-mapping.yaml
role_mappings:
  # Direct mapping from IdP group to application role
  "Okta: BI-Admins" -> "admin"
  "Okta: BI-Users" -> "user"
  
  # Regex pattern for broader matching
  "AD: IT-Staff-.*" -> "it_support"
  
  # Default role if no other mappings match
  default_role: "guest"
  
  # Defines role inheritance
  role_hierarchy:
    admin: ["user", "guest"]
    user: ["guest"]
\end{lstlisting}
\end{tcolorbox}
\caption{An example of a decoupled role mapping configuration, demonstrating direct, pattern-based, and default role assignments, which provides greater flexibility than hardcoded logic.}
\label{fig:rbac_config}
\end{figure}

\begin{figure}[h]
\begin{tcolorbox}[colback=backcolour,colframe=black!75!white,title={Main Orchestration Logic in the Auth Shim}]
\begin{lstlisting}[language=Python]
def process_saml_login(saml_response):
    """Main function to handle the complete login flow."""
    # 1. Validate the SAML assertion
    saml_auth = validate_saml(saml_response)
    user_info = get_user_info_from_saml(saml_auth)
    
    # 2. Get the required admin token for the app's API
    admin_token = os.environ.get('APP_ADMIN_TOKEN')
    
    # 3. Find, create, or reactivate the user
    user_id = find_or_create_user(user_info, admin_token)
    
    # 4. Synchronize user roles based on IAM groups
    sync_user_roles(user_id, user_info['groups'], admin_token)
    
    # 5. Create a new session for the user in the app
    session_token = create_app_session(user_info['email'])
    
    # 6. Return the session token to set in the browser
    return session_token
\end{lstlisting}
\end{tcolorbox}
\caption{High-level orchestration function showing the step-by-step logic executed by the Auth Shim upon receiving a valid SAML response.}
\label{fig:main_orchestration_code}
\end{figure}

\subsection{Process and User Journey Diagrams}

\begin{figure*}[t!]
    \centering
    \includegraphics[width=0.9\textwidth]{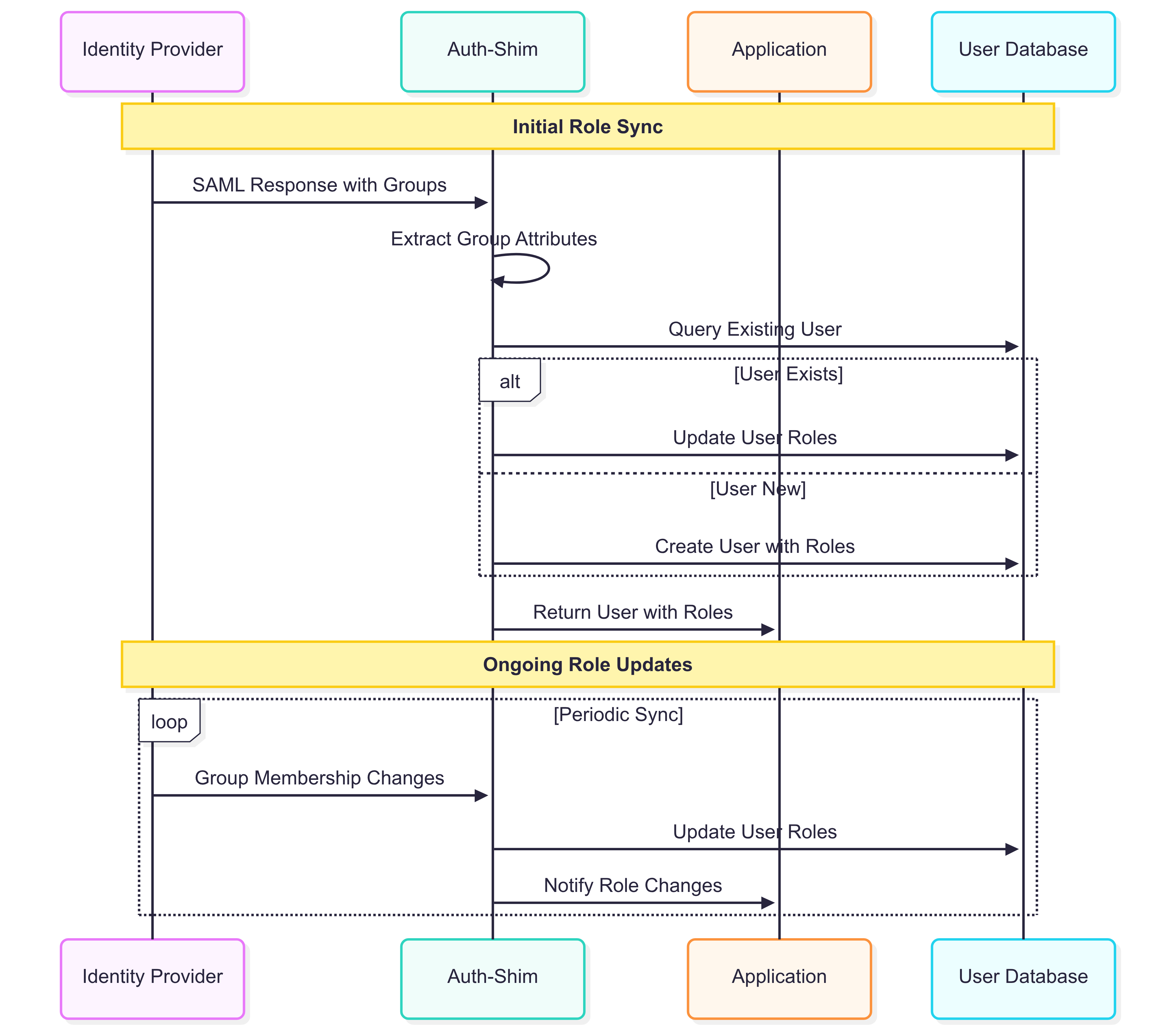}
    \caption{The step-by-step logic of the Just-in-Time role synchronization process, from SAML attribute extraction to final session generation.}
    \label{fig:role_sync_flowchart}
\end{figure*}

\begin{figure*}[t!]
    \centering
    \includegraphics[width=0.8\textwidth]{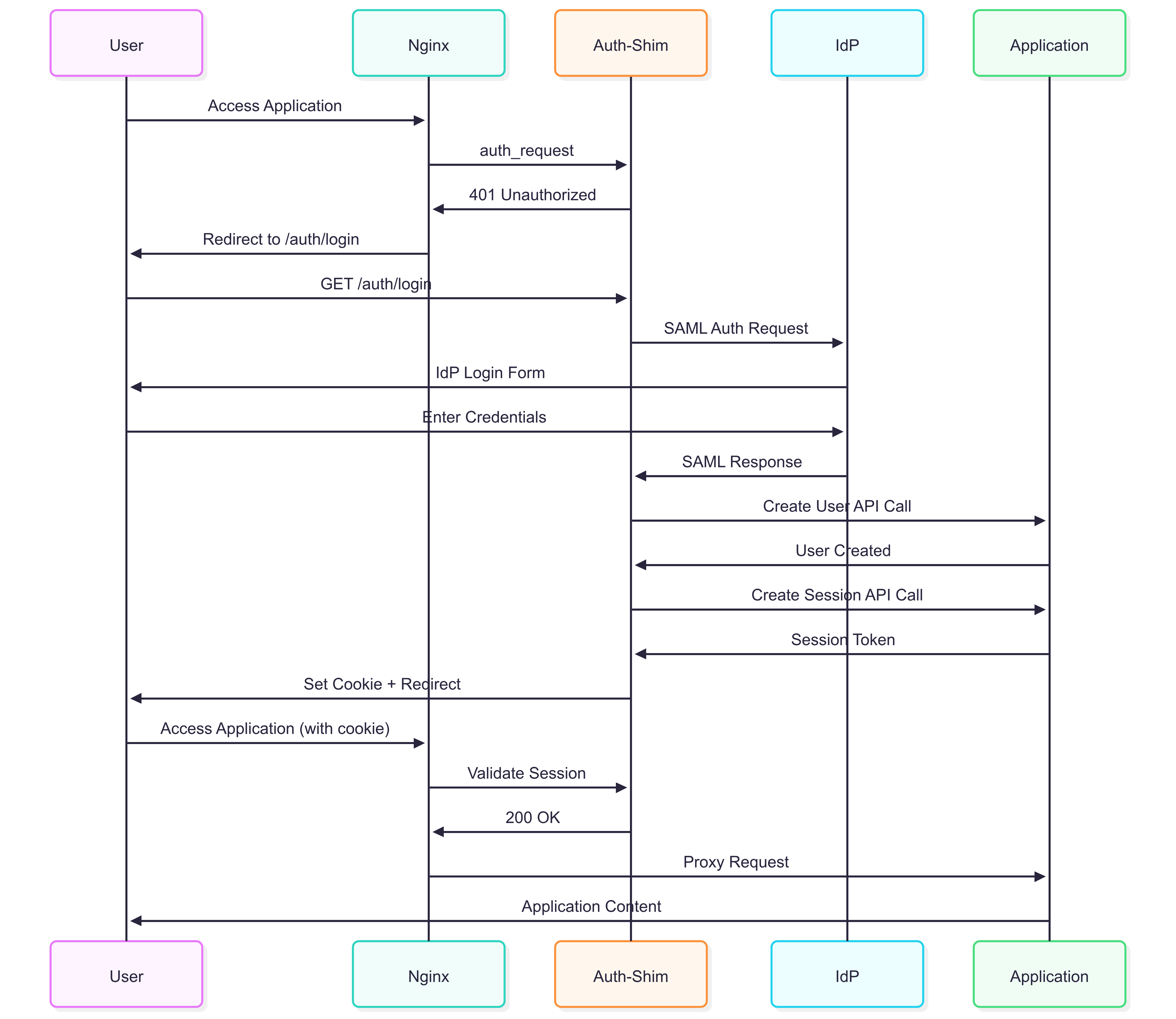}
    \caption{Sequence diagram for a first-time user login, demonstrating automated user provisioning.}
    \label{fig:journey_first_time}
    \vspace{4em} 
    \includegraphics[width=0.8\textwidth]{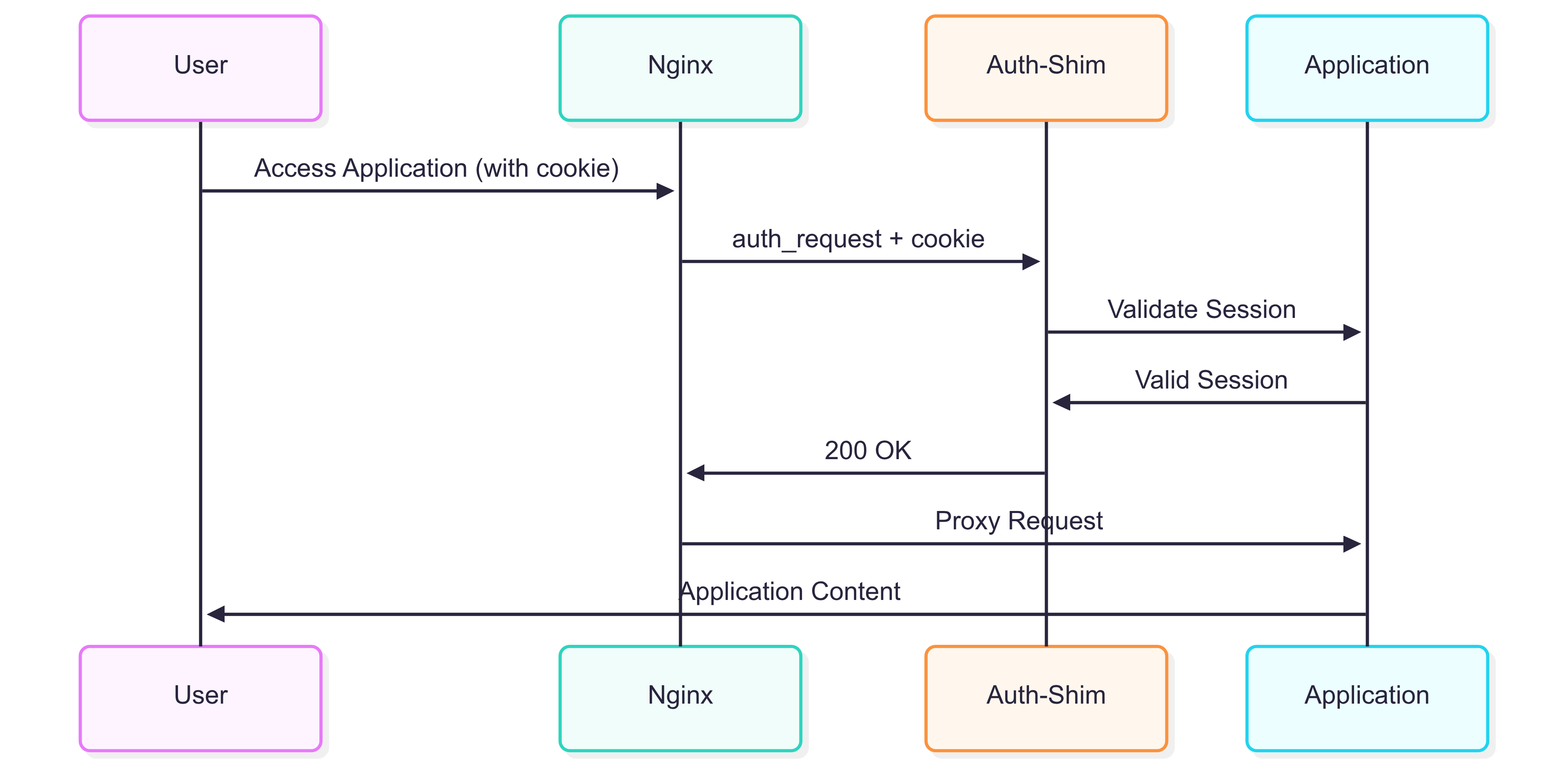}
    \caption{Sequence diagram for a returning user with a valid session, showing the fast-path authentication check.}
    \label{fig:journey_returning}
\end{figure*}


\begin{thebibliography}{99}

\bibitem{report:opensourcestate}
OpenLogic by Perforce.
\newblock \emph{2023 State of Open Source Report}.
\newblock 2023. Available at: \url{https://www.openlogic.com/resources/2023-state-of-open-source-report} (Accessed: 2023-11-15).

\bibitem{paper:zerotrust}
John Kindervag.
\newblock No More Chewy Centers: Introducing The Zero Trust Model Of Information Security.
\newblock \emph{Forrester Research}, Sep 2010.

\bibitem{paper:beyondcorp}
Rory Ward and Betsy Beyer.
\newblock BeyondCorp: A New Approach to Enterprise Security.
\newblock \emph{\;login: The USENIX Magazine}, 39(6), Dec 2014.

\bibitem{paper:identityfed}
P.~A. Karger and J.~S. Park.
\newblock Identity Federation: Issues and Architectures.
\newblock In \emph{2006 IEEE Security and Privacy Workshops}, pp. 11--11, 2006.
doi:10.1109/SPW.2006.19.

\bibitem{book:designpatterns}
Erich Gamma, Richard Helm, Ralph Johnson, and John Vlissides.
\newblock \emph{Design Patterns: Elements of Reusable Object-Oriented Software}.
\newblock Addison-Wesley Professional, 1994. ISBN: 0201633612.

\bibitem{paper:legacyIAM}
Christoph Arndt.
\newblock Secure and Scalable Legacy IAM Integration.
\newblock \emph{IEEE Software}, 39(1):89--94, 2022.
doi:10.1109/MS.2021.3119102.

\bibitem{paper:apigatewaysec}
C.~Pahl, A.~Jamshidi, and O.~Zimmermann.
\newblock Architectural Patterns for Secure and Trustworthy API Gateways.
\newblock In \emph{2018 IEEE International Conference on Software Architecture (ICSA)}, pp. 123--12307, 2018.
doi:10.1109/ICSA.2018.00021.

\bibitem{paper:microproxyrbac}
S.~Hassan and G.~Russello.
\newblock A Micro-proxy for Enforcing Attribute-Based Access Control in Microservices.
\newblock In \emph{Proceedings of the 13th ACM Symposium on Access Control Models and Technologies}, pp. 191--200, 2008.
doi:10.1145/1377836.1377865.

\bibitem{paper:ssosec}
A.~Armando, R.~Carbone, L.~Compagna, J.~Cuellar, and G.~Pellegrino.
\newblock On the Security of Single Sign-On Protocols in the Wild.
\newblock In \emph{2015 IEEE Symposium on Security and Privacy (SP)}, pp. 611--628, 2015.
doi:10.1109/SP.2015.43.

\bibitem{pomerium_docs}
Pomerium.
\newblock \emph{Pomerium Documentation}.
\newblock 2024. Available at: \url{https://www.pomerium.com/docs/} (Accessed July 2025).

\bibitem{ory_oathkeeper}
ORY.
\newblock \emph{ORY Oathkeeper Docs}.
\newblock 2024. Available at: \url{https://www.ory.sh/oathkeeper/} (Accessed July 2025).

\bibitem{plugin_limitations}
OpenLogic.
\newblock Challenges with Commercial SSO Plugins.
\newblock 2023. Available at: \url{https://www.openlogic.com/blog/sso-plugin-limitations} (Accessed July 2025).

\bibitem{oauth2proxy_docs}
OAuth2 Proxy.
\newblock \emph{OAuth2 Proxy Documentation}.
\newblock 2024. Available at: \url{https://oauth2-proxy.github.io/oauth2-proxy/} (Accessed July 2025).

\bibitem{tech:docker}
Docker, Inc.
\newblock Docker Documentation.
\newblock 2023. Available at: \url{https://docs.docker.com/} (Accessed: 2023-11-15).

\bibitem{tech:stride}
Microsoft.
\newblock The STRIDE Threat Model.
\newblock 2023. Available at: \url{https://learn.microsoft.com/en-us/azure/security/develop/threat-modeling-tool-threats} (Accessed: 2023-11-16).

\end{thebibliography}
\end{document}